\documentclass[12pt]{article}%
\usepackage{amsfonts}
\usepackage{bm}
\usepackage{ulem}
\usepackage{cite}
\usepackage{mathrsfs}
\usepackage{amsmath}
\usepackage{graphicx}
\usepackage{hyperref}
\usepackage{verbatim}
\usepackage{color}
\usepackage{enumerate}
\usepackage{amsfonts}
\usepackage{amssymb}
\usepackage{authblk}
\usepackage[hang,flushmargin]{footmisc}
\usepackage{lipsum}
\usepackage[font=footnotesize]{caption}
\usepackage{soul}
\usepackage[utf8]{} 
\usepackage{empheq}

\csname @addtoreset\endcsname{equation}{section}
 \newcommand{\badat}{\begin{alignedat}}
 \newcommand{\eadat}{\end{alignedat}}

\def\be{\begin{eqnarray}}
\def\ee{\end{eqnarray}}
\def\beann{\begin{eqnarray*}}
\def\eeann{\end{eqnarray*}}
\def\beq{\begin{equation}}
\def\eeq{\end{equation}}
\def\ba{\begin{array}}
\def\ea{\end{array}}
\def\ben{\begin{enumerate}}
\def\een{\end{enumerate}}
\def\bea{\begin{eqnarray}}
\def\eea{\end{eqnarray}}

\def\5{\bar }
\def\6{\partial }
\def\7{\hat }
\def\4{\tilde }

\def\cA{\mathcal{A}}

\def\cH{\mathcal{H}}
\def\cI{\mathcal{I}}
\def\cJ{\mathcal{J}}

\def\cP{\mathcal{P}}

\newcommand{\mean}[1]{\left\langle #1 \right\rangle} 

\newcommand{\comm}[2]{\left[#1,#2\right]}

\renewcommand{\d}{\partial}

\renewcommand{\tilde}{\widetilde}
\renewcommand{\hat}{\widehat}

\begin{document}

\title{\vspace{-70pt} \Large{\sc  DC conductance and memory  in 3D gravity}\vspace{10pt}}
\author[a]{\normalsize{M\'arcela C\'ardenas}\footnote{\href{mailto:marcela.cardenas@usach.cl}{marcela.cardenas@usach.cl}}}
\author[b]{\normalsize{Hern\'an A. Gonz\'alez}
\footnote{\href{mailto:hernan.gonzalez@uai.cl}{hernan.gonzalez@uai.cl}}}
\author[a]{\normalsize{Kristiansen Lara}
\footnote{\href{mailto:kristiansen.lara@usach.cl}{kristiansen.lara@usach.cl}}}
\author[a,c]{\normalsize{Miguel Pino}\footnote{\href{mailto:miguel.pino.r@usach.cl}{miguel.pino.r@usach.cl}}}

\affil[a]{\footnotesize\textit{Departamento de F\'isica, Universidad de Santiago de Chile, Avenida Víctor Jara 3493, Santiago, Chile.}}
\affil[b]{\footnotesize\textit{Facultad de Artes Liberales, Universidad Adolfo Ib\'a\~nez, Diagonal Las Torres 2640, Santiago, Chile.}}
\affil[c]{\footnotesize\textit{Center for Interdisciplinary Research in Astrophysics and Space Exploration (CIRAS),
Universidad de Santiago de Chile, Av. Libertador Bernardo O’Higgins 3363, Santiago, Chile.}}

\date{}

\maketitle
\thispagestyle{empty}
\begin{abstract}
Transport properties are investigated in the two-dimensional dual dynamics of AdS$_{3}$ gravity. By providing boundary conditions that deform the ADM lapse and shift functions, we construct a lower dimensional model comprising two copies of chiral boson excitations with anisotropic scaling symmetry. Using bosonization, an electric current is identified. By means of the Kubo formula, we find a DC conductance depending on the level of the theory and the dynamical exponents. The bulk realization of the linear response is related to a type of gravitational memory emerging in the context of near-horizon boundary conditions. The  process is adiabatic and represents a permanent spacetime deformation parametrized by anisotropic chiral bosons through a large gauge transformation. 
\end{abstract}

\newpage

\begin{small}
{\addtolength{\parskip}{-2pt}
 \tableofcontents}
\end{small}
\thispagestyle{empty}
\newpage

\section{Introduction}

 The Chern-Simons description of Einstein gravity in three dimensions\cite{Achucarro:1987vz,Witten:1988hc} stresses the fact that properties of the spacetime are specified by the topology of the manifold and the presence of boundaries.  In other words, the dynamical content of 3D gravity is determined after globally characterizing the manifold and  providing suitable boundary conditions.  From a holographic point of view, the asymptotic  symmetries respecting these configurations, give rise to the global symmetries of the dual theory.
 
 Recently, the specification of field-dependent Lagrange multipliers at infinity has permitted to propose novel boundary dynamics, via imposition of integrability conditions consistent with the gauge symmetries \cite{Henneaux:2013dra, Perez:2016vqo, Ojeda:2019xih, Cardenas:2021vwo}. In the case of asymptotically AdS$_3$ spaces, this procedure deforms the Hamiltonian of the boundary theory, describing mechanical systems with symmetries other than conformal. 

The holographic representation of three-dimensional gravity finds interesting applications in the spirit of the AdS/CMT correspondence  (for a review on the subject see e.g. \cite{Hartnoll:2009sz,McGreevy:2009xe}), where observables sensible for condensed matter systems can be recognized from the symmetries of the gravitational description. 

In this work, we would like to compute transport coefficients for three-dimensional gravity endowed with a particular choice of the Lagrange multipliers. This specification makes contact with Lifshitz symmetry, that has shown to be relevant in condensed matter systems \cite{sachdev_2011}. We construct the corresponding dual theory which describes anisotropic chiral boson excitations \cite{Fuentealba:2019oty}, admitting a conservation law based on a residual $U(1)$ symmetry.  The resulting conserved current can be reinterpreted under the light of bosonization\footnote{See \cite{vonDelft:1998pk} for a discussion from the point of view of condensed matter systems.}, permitting to identify an electronic charge density and its corresponding electric current.
By means of the Kubo formula, we obtain the linear response associated to the charge-current correlator. This computation has been previously carried out in the relativistic chiral boson case by Kane and Fisher \cite{kane1996edge}. In the DC limit, this correlator yields an Ohm's law when the physical setup is characterized by a source representing a two-terminal reservoir. Here, we present a generalization of \cite{kane1996edge} to the case of anisotropic chiral states. 
Noteworthy, the specification of our boundary conditions, contained in the spatial and temporal components of the  Chern-Simons fields, is shown to play a precise role in the functional expression of the DC conductance. 


The gravitational realization of the transport phenomena is achieved by considering gauge fields  describing the near horizon region of a black hole  \cite{Afshar:2016wfy,Afshar:2016kjj}. This gives a complementary bulk description of the charge density and electric current operators, that in this case, control the horizon area and the angular velocity measured by an observer hovering just outside the horizon region. We consider that these observables are subjected to the same external source employed to calculate the edge-state transport coefficient.

The bulk counterpart of the boundary linear response has also an appealing relation with memory effects\cite{Zeldovich:1974gvh,Strominger:2014pwa}. 
In spite that 3D Einstein gravity is devoid of local gravitational excitations, the spacetime deformation due to the external source can be retrieved in terms of an improper gauge transformation. Remarkably, this transformation encodes the DC transport coefficient and shows a correspondence with the boundary computation.

This paper is organized as follows: In section \ref{Section1}, we revisit the construction of the boundary dynamics from a Chern-Simons action. We provide boundary conditions making contact with the model proposed in \cite{Fuentealba:2019oty}. 
More precisely, the dynamics turns out to be described by two copies of anisotropic chiral boson. 
In section \ref{Section2}, we study transport properties of the anisotropic chiral boson. Using bosonization, we identify an electric charge and current operator. We obtain the expectation value of these operators in frequency space. In the DC limit, we show that the two-terminal conductance receives contributions from the coupling constants characterizing both chiralities. In section \ref{Section3}, we rederive this conductance from the perspective of an observer in the near-horizon region and  discuss 
a gravitational memory interpretation. In Section \ref{conclusion}, we conclude presenting the main ideas of work and we propose some future prospects.

\section{Anisotropic chiral bosons from a Hamiltonian reduction} \label{Section1}

This section reviews the construction of the boundary dynamics in three-dimensional gravity using the Chern-Simons formulation. By imposing appropriate boundary conditions, we make contact with a recently introduced model dubbed {\it anisotropic chiral boson} \cite{Fuentealba:2019oty}. The derivation generalizes the seminal example presented in \cite{Coussaert:1995zp}, where Brown-Henneaux boundary conditions give rise to a theory of two decoupled chiral bosons.

As it will be shown  below, the boundary dynamics will be completely defined by the specification of a symplectic structure and a Hamiltonian obtained from a well-posed variational principle.

\subsection{Kinetic term}

The starting point of the construction considers the action principle given by the difference of two Chern-Simons functionals $I_{CS}[A^+]-I_{CS }[A^-]$, where the connections $A^\pm$ are $\mathfrak{sl}(2,\mathbb{R})$ valued one-forms. The spacetime has the topology $\mathbb{R}  \times \Sigma $, where the real line is parameterized by time $t$, while $\Sigma$ is a disc endowed with coordinates $r>0$ and $0\leq x<2 \pi l$, where $l$ stands for the AdS radius. The level is $K=\frac{l}{4G}$ where $G$ is the Newton constant. See appendix \ref{Appendix A}, for details on conventions and explicit formulas. 

The boundary symplectic structure can be obtained by solving the bulk constraints and replacing them back in the three-dimensional action.  Since the theory is devoid of local degrees of freedom, the Chern-Simons kinetic term becomes a total derivative that can be expressed as a local two-dimensional functional \cite{Elitzur:1989nr,Coussaert:1995zp}. 

Here, an alternative path is taken for constructing the reduced kinetic term of the model. In doing so, it is necessary to study the symmetries allowed by the field configurations  satisfying a specific boundary condition at constant time slice. In the Chern-Simons formulation, asymptotic conditions are encoded in the spatial component of the gauge field
\be 
A^\pm_{x}=a^\pm_x+\cdots\,,
\ee
where the ellipsis represents subleading terms close to the boundary, capturing the radial dependence. The set of gauge transformations $\varepsilon^\pm$ preserving the asymptotic form of $A^\pm_x$ are global as long as they produce non-vanishing surfaces charges \cite{Regge:1974zd}. In that case, they are given by
\be
\label{QQ}
\delta Q(\varepsilon^+, \varepsilon^-)=-\frac{K}{2\pi} \int dx\,\langle \varepsilon^+ \delta a^+_x - \varepsilon^- \delta a^-_x\rangle\,.
\ee
Without imposing any further restriction on $a^\pm_x$, the asymptotic symmetries are generated by the $\mathfrak{sl}(2,\mathbb{R})$ current algebra \cite{Banados:1998gg,Grumiller:2016pqb}. We consider the simplest set of connections within the $\mathfrak{sl}(2,\mathbb{R})$ family spanned by only one generator \cite{Afshar:2016kjj}
\be \label{ax}
a^{\pm}_x=\pm \frac{4\pi}{K}\cJ_{\pm}(x,t) \; L_0\,.
\ee
Gauge parameters respecting the above boundary conditions are of the form $\varepsilon^{\pm}=\varepsilon^{\pm}(x,t)L_0$. The corresponding charge densities are  $\cJ_\pm$, giving rise to $\mathfrak{u}(1)$ current algebras under Dirac brackets
\be
\label{chargesalg}
\left\{\cJ_{\pm}(x),\cJ_{\pm}(x')\right\}_{\rm D. B.}= \pm \frac{K}{4\pi}\d_x \delta(x-x')\,.
\ee
The latter algebra  can be used to define the canonical variables at the boundary. In fact, as shown in \cite{Floreanini:1987as}, the kinetic term producing commutation relations \eqref{chargesalg} is 
\be
I_{\rm kinetic}[\cJ_{+},\cJ_{-}]= \frac{2\pi}{K} \int dx dy dt \;\epsilon (x-y) \left[ \cJ_{+}(x) \d_t\cJ_{+}(y) - \cJ_{-}(x) \d_t\cJ_{-}(y) \right]\,,
\ee
where $\epsilon(x)$ is the inverse of operator $\d_x$. In order to obtain a local description, we use fields $\phi_\pm$ that are defined by
\be
\label{eq:charges}
\cJ_{\pm}=\pm\frac{K}{4\pi} \d_{x}\phi_{\pm}\,,
\ee
and the kinetic term becomes
\be
I_{\rm kinetic}[\phi_+,\phi_-]= \frac{K}{8\pi} \int dx  dt \; \left[ \d_x \phi_{+} \d_t\phi_{+} - \d_x \phi_{-} \d_t\phi_{-} \right]\,.
\ee

\subsection{Boundary Hamiltonian}

To define the time evolution generator of the system, we consider boundary charges conjugated to the gauge transformations $\varepsilon^\pm=-a^\pm_t$, where $a^\pm_t$ is the leading term of the temporal component $A^\pm_t$. By using \eqref{QQ}, we find that
\be
\label{aqui}
\delta H=\frac{K}{2\pi}\int dx \langle a^+_t \, \delta a^+_x - a^-_t \, \delta a^-_x \rangle\,.
\ee
A local expression for the Hamiltonian can be obtained provided the leading terms $a^\pm_t$ fulfill a condition of the form
\be \label{at1}
a^\pm_t=\frac{\delta \cH_\pm(a^\pm_x)}{\delta a^\pm_x}\,,
\ee
where  $\cH_\pm(a^\pm_x)$ is local functional of $a^\pm_x$. However, the choice of the latter functional is not completely arbitrary since time evolution of the system must  also preserve the asymptotic behavior of $A^\pm_x$. To ensure this condition, one then chooses $a^\pm_t$ to have the same functional form than the allowed gauge parameters $\varepsilon^{\pm}$ \cite{Henneaux:2013dra}.
For instance, in the case of Brown-Henneaux boundary conditions without chemical potentials \cite{Coussaert:1995zp}, the choice $a^\pm_t=\pm a^\pm_x$ singles out a Hamiltonian that is quadratic in $a^\pm_x$ and produces modes evolving in the $t \pm x$ directions. Another possible choice has been considered in \cite{Perez:2016vqo,Cardenas:2021vwo} in connection with integrable systems. In that case, the selected functional $\cH_\pm(a^\pm_x)$ is given by AKNS charges and the corresponding Hamiltonian evolution of the modes encompasses the AKNS hierarchy \cite{Ablowitz:1974ry}.  

With the purpose of studying time-dependent perturbations in a family of 2D field theories, we focus on the particular  class   
\be
\label{integra}
a^\pm_t=  a_{0}^\pm \pm v (-1)^{\frac{z-1}{2}} \d^{z-1}_x a^\pm_x \,, 
\ee
where $a_{0}^\pm$ are arbitrary functions of $t$ without variation. This term has been added in order to guarantee the existence of a non-degenerate background metric field  \footnote{ The metric field  can be constructed from
$
g_{\mu \nu}=\frac{l^2}{2} \mean{(A^+_\mu-A^-_\mu),(A^+_\nu-A^-_\nu)}\,.
$
}. It is important to stress that the deformation generated by \eqref{integra} produces a field-dependent asymptotic behavior of the lapse and shift functions in the ADM metric formulation \cite{Perez:2016vqo,Cardenas:2021vwo}.


The dynamical exponent $z$  is an odd number, that can take different values for each copy, while $v>0$ is a coupling constant with dimensions of $[{\rm length}]^{z-1}$. 
In the upcoming section, we will use distinct dynamical exponents $z_{\pm}$ on each Chern-Simons copy. Notice that only the particular choice $z_+=z_-$ confers Lifshitz symmetry to the boundary dynamics of the gravitational model. In particular, the Brown-Henneaux case is contained in the choice $z_\pm=1$, where the scaling symmetry becomes isotropic.

The Hamiltonian \eqref{aqui} can be readily integrated using the condition \eqref{integra}. Additionally, when implementing $a^\pm_x=\d_x \phi_\pm L_0$, its expression depends on the local fields $\phi_\pm$, yielding
\be
\label{H0}
H=H_0[\phi_+]+H_0[\phi_-]\,,
\ee
where  
\begin{align}
\label{H}
H_0[\phi]= \frac{Kv}{8\pi} \int d x \,  \left( \d^{\frac{z_\pm+1}{2}}_x \phi  \right) ^2 \,.
\end{align}
 
 \subsection{Anisotropic chiral boson spectrum}
 
Let us describe the action for each chirality
\begin{align}
\label{e0}
I[\phi_\pm]=\pm \frac{K}{8\pi} \int dx  dt \; \left[ \d_x \phi_{\pm} \d_t\phi_{\pm} \mp  v \left( \d^{\frac{z+1}{2}}_x \phi_\pm  \right) ^2\right].
\end{align}
For sake of simplicity, here we only consider one odd dynamical exponent $z$. The theory realizes anisotropic scaling symmetry $t \to \lambda^z t$ and $x \to \lambda  x$. Excitations are chiral in the sense that parity transformation ($x \to -x$) swaps the dynamics of $\phi_\pm$ into $\phi_\mp$. 

The equations of motion  corresponds to
\be\label{bosoneq}
\d_x \left(\d_t \phi_{\pm}\pm (-1)^{\frac{z+1}{2}} v \, \d_{x}^z \phi_{\pm}\right)=0\,.
\ee
Since this is a linear equation, solutions are given by plane wave modes 
\be
\label{eqphi}
\phi_{\pm}= \exp[i(kx \pm \omega_k t)]\,, \quad \omega_k= v k^z\,.
\ee
Furthermore, due to parameterization \eqref{eq:charges} the system inherits  zero modes responsible for the gauge symmetry $\phi_{\pm}\to\phi_{\pm}+f_{\pm}(t)$. In what follows, we consider these to be gauge fixed and therefore they will not be present in the subsequent analysis.

Another important feature is that the model \eqref{e0} exhibits a $\mathfrak{u}(1)$ current symmetry on each sector. Under the infinitesimal transformation $\delta \phi_{\pm}= \eta^{\pm}_k\exp[i(kx \pm \omega_{k} t)]$, with $\eta^{\pm}_k \ll 1$, the Noether theorem yields the expected currents \eqref{eq:charges}.

For the forthcoming analysis, we describe the fields in terms of their Fourier modes. 
In the remainder of the paper, we will be concerned with local properties of the theory, hence $x$ is regarded as a coordinate defined on the whole real line. This allows to expand the field operator as
\begin{align}
\phi_{\pm}(x,t)= \int^{+\infty}_{-\infty} \frac{d k}{ k} e^{i(kx \pm \omega_k t)} b_{\pm,k}\,,
\end{align}
where the condition $b_{\pm,k}^{\dagger}=-b_{\pm,-k}$ ensures hermiticity of $\phi_{\pm}$. 

The Dirac bracket can be read off from the kinetic term in \eqref{e0}. The quantum version yields the equal time commutation relation\footnote{In the quantum theory we use the rule $\left[ ,\right] \to i\{ ,\}_{\rm DB}$.}
\begin{align} \label{boson comrel}
\left[ \phi_{\pm} (t,x), \partial_{x'} \phi_{\pm}(t,x') \right]=  \pm \frac{4\pi}{K} i \delta(x-x')\,.
\end{align}
In terms of the $b_{\pm,k}$ operators, they become
\begin{align}
\left[ b_{\pm,k}, b_{\pm,k'}\right]=  \pm \frac{2}{K} k \delta(k+k')\,.
\end{align}

\section{Linear response and the DC limit}
\label{Section2}

This section explores the dynamical evolution of the boundary fields $\phi_\pm$ driven by a perturbation to the free  Hamiltonian of the boundary theory. At the level of the Lagrangian density \eqref{H0}, such deformation  is chosen as,
\be\label{DeltaHphi}
\Delta H = \int\, dx \, \mu(x,t) \, \left[\cJ_{+}(x,t)+\cJ_{-}(x,t)\right]\,,
\ee
where $\cJ_\pm$ are the Noether currents expressed in terms of the fields $\phi_\pm$ by \eqref{eq:charges} and $\mu$ is an external source localized in time, that will model an evolution process between vacuum solutions of \eqref{bosoneq}. As shown below, this  will allow us to calculate transport coefficients by means of the Kubo formula.

\subsection{Response function}

We start by identifying a conservation law using the charges expressed in terms of the chiral boson \eqref{eq:charges}
\be
\label{eq:cons}
\d_{t} \cJ_{\pm}+ \d_x \cI_{\pm}=0\,,
\ee
implying that $\cI_{\pm}$ is 
\begin{align}
\label{eq:charges2}
\cI_{\pm}=\mp \frac{K}{4\pi} \d_t \phi_{\pm}\,.
\end{align}
Equation \eqref{eq:cons} associates $\cJ_\pm$ with a density and  $\cI_\pm$ with a current. 
In the context of bosonization, they can be understood as carriers of fermionic degrees of freedom, so that $\cJ_{\pm}$ is associated to an electric charge density and $\cI_\pm$ corresponds to an electric current flowing along the boundary. For details  on the precise relation between $\cJ_{\pm}$ and fermionic operators see appendix \ref{Appendix BB}. 

We wish to evaluate the linear response of the net current  $\cI_{\rm tot}\equiv \cI_{+}+\cI_{-} $,  induced by the deformation \eqref{DeltaHphi}, which is interpreted as an electric potential energy. In this sense, the source $\mu$ produces a voltage creating a current in a one-dimensional conductor. In the next subsection, we will be interested in a source $\mu$ creating a flow between two reservoirs. This has been interpreted by \cite{kane1996edge} as edge states flowing at the boundary of a two-dimensional material. 

We analyze the expectation value of the variation of the current using the Kubo formula in frequency space (we refer the reader to appendix \ref{Appendix C} for conventions)
\begin{align}
\label{qqq}
\delta \langle \tilde{\cI}_{\pm}(\omega,x) \rangle &=  \int \,dx'\, \tilde{\mu}(x',\omega)  \tilde{\chi}^{\pm}_{\cI,\cJ}(\omega; x,x')\,,
\end{align}
where the susceptibility $\tilde{\chi}^{\pm}_{\cI,\cJ}$ is
\begin{align}
\label{sss0}
\tilde{\chi}^{\pm}_{\cI,\cJ}(\omega; x,x')= - i \int dt\, \Theta(t) e^{i\omega t} \mean{ \left[ \cI_{\pm}(x,t), \cJ_{\pm}(x',0)\right]} \,.
 \end{align}
After using \eqref{boson comrel} and its Fourier mode expansion,  the suceptibility can be expressed as
\begin{align}
\label{sss}
\tilde{\chi}^{\pm}_{\cI,\cJ}(\omega; x,x')=\frac{K}{4\pi} \left[\pm \delta(x-x') \mp (\omega+i\epsilon) F^{\pm}(x-x',\omega) \right]\,.
\end{align}
The functions $F^{\pm}$ have been obtained after expressing the Heaviside function $\Theta(t)$ as an integral in frequency space, inducing the shift  $\omega \to \omega+i\epsilon$. They become
\be
(\omega+i\epsilon) F^{\pm}(y,\omega)= i \Theta(-y) \sum_{k^{\pm}_p  \in {\rm Im}_{<} }\frac{ \omega(k^{\pm}_p)  e^{i k^{\pm}_p y}}{ \omega'(k^{\pm}_p)} - i \Theta(y) \sum_{k^\pm_p \in {\rm Im}_{>}} \frac{\omega(k^{\pm}_p) e^{i k^{\pm}_p y} }{\omega'(k^{\pm}_p)}\,,
\ee
where  ${\rm Im}_{\gtrless}$ correspond to the upper and lower complex plane.  The values $k^{\pm}_p=k^{\pm}_p(\omega)$ are solutions to the equation $\omega+i\epsilon \pm \omega(k^{\pm}_p)=0$.

\subsection{Two-terminal conductance}

The situation of interest is modeled by an  external source that generates a flow between two reservoirs  located at positions $x=x_L$ and $x=x_R$ \cite{kane1996edge}, that perturbs the system for a very short time interval. 
In order to reproduce these features we use
\be
\label{mu}
\tilde{\mu}(x,\omega)= V_L \,\Theta(x_L-x)+V_R\,\Theta(x-x_R)\,.
\ee
Notice that time localization of the perturbation is ensured as the inverse Fourier transform of $\tilde{\mu}$ yields a temporal Dirac delta. 

In the presence of the source $\tilde{\mu}$, the  variation of the electric current on each sector becomes
\begin{multline}
\label{Ipm}
\delta \langle  \tilde{\cI}_{\pm} \rangle = \mp \frac{K}{4\pi}\Bigg\{V_L \left[\Theta(x-x_L) \Delta^{\pm}_{>}(x-x_L)   -\Theta(x_L-x)\Delta^{\pm}_{<}(x-x_L) \right] \\ -V_R \left[\Theta(x-x_R) \Delta^{\pm}_{>}(x-x_R) 
- \Theta(x_{R}-x)\Delta^{\pm}_{<}(x-x_R) \right]   \Bigg\}\,,
\end{multline}
where
\be
\Delta^{\pm}_{\gtrless}(x) =\frac{1}{z^{\pm}}\sum_{k^\pm_p \in {\rm Im}_{\gtrless}} e^{ik^{\pm}_p x}\,.
\ee
As we previously mentioned, we have considered a Hamiltonian \eqref{H} with different dynamical exponents on each sector, given by $z_+$ and $z_-$ respectively.

In the zero frequency limit, we obtain the DC current contribution. Noticing that $k^{\pm}_p(\omega \to 0)=0$, we see that $\Delta^{\pm}_{\gtrless}$ becomes a constant that essentially counts the number of zeros of the equation 
\be
\omega+i\epsilon \pm v (k^{\pm}_p)^{z^{\pm}}=0\,, \nonumber
\ee 
on the upper/lower complex plane. Causality, represented by the $i\epsilon$-prescription, implies that the $\pm$ sector has $\tfrac{z^{\pm}\mp1}{2}$ zeros on the upper plane and  $\tfrac{z^{\pm}\pm1}{2}$ zeros on the lower one. Therefore, functions  $\Delta^{\pm}_{\gtrless}$ in DC limit read
\be
\lim_{\omega \to 0}\Delta^{\pm}_{>}=\frac{z^{\pm}\mp1}{2z^{\pm}}\,, \quad 
\lim_{\omega \to 0} \Delta^{\pm}_{<}=\frac{z^{\pm}\pm1}{2z^{\pm}}\,. 
\ee
Using these expressions, the variation of the net current $\langle \tilde{\cI}_{\rm tot}\rangle$  satisfies  Ohm's law. Indeed, by denoting  the voltage as $\Delta V= V_L-V_R$, we find that the two-terminal conductance $\sigma$ \cite{kane1996edge} is given by 
\be
\label{conductivity}
\quad \sigma \equiv \frac{\delta \langle \tilde{\cI}_{\rm tot}  \rangle}{\Delta V}=\frac{K}{8\pi}\left( \frac{1}{z_+}+\frac{1}{z_-}\right)\,.
\ee
The above expression have two contributions that are worth commenting on. Firstly, there is a factor coming from the level $K=\frac{l}{4G}$ associated to the infinite-dimensional $\mathfrak{u}(1)$ current symmetry \eqref{chargesalg}. This symmetry is universal in the sense that is controlled by the residual transformation of the $a^{\pm}_x$. Lastly, the conductivity is also modified by the dynamical exponents of the theory $(z_+,z_-)$  that are associated to the higher derivative Hamiltonians \eqref{H}. This contribution is model dependent as it emerges from the choice of $a^{\pm}_t$.

\section{The gravitational side}\label{Section3}

We wish to understand gravitational implications associated to the DC response observed on the boundary description. For that purpose, we construct the bulk observables that were identified with electric properties in the preceding calculation and probe the effect of the perturbation on a near horizon metric field. We will show that from a bulk perspective, the source $\mu$  induces a permanent deformation on the spacetime that can be interpreted with a type of gravitational memory effect.

We start by defining the Chern-Simons fields $A^{\pm}$ for the three-dimensional theory, whose field equations are modified by external sources as follows\footnote{The action principle giving rise to these field equations is discussed in appendix \ref{Appendix A}. The orientation of the manifold is fixed by $\epsilon^{rxt}=1$.}

\be
\label{FJ}
\frac{K}{4\pi}\epsilon^{\alpha \mu \nu}F^{\pm}_{\mu \nu}=J^{\alpha}_{\pm}\,.
\ee
The gauge connections are related with their asymptotic values $a^{\pm}$ through the decomposition,
\be
\label{Aa}
A^{\pm}=b_{\pm}^{-1} a^{\pm} b_{\pm} + b_{\pm}^{-1} d b_{\pm}\,.
\ee
Here, $b_{\pm}$ is an SL$(2,\mathbb{R})$ group element depending on the radial coordinate $r$. The longitudinal gauge connections are $a_{\pm}=a^{\pm}_x dx + a^{\pm}_t dt$, with $a^{\pm}_x$ given by \eqref{ax}, while the temporal component is
\be
\label{at}
a^{\pm}_t\equiv\mp\frac{4\pi}{K}\cI_{\pm}\, L_0\,.
\ee 
The source is defined through a current pointing in the radial direction
\be
\label{JJ}
J^r_\pm =-\frac{K}{2\pi}\d_{x} \mu  \,\, b_{\pm}^{-1}L_{0}b_{\pm} \,, \quad J^t_\pm =J^x_\pm =0\,,
\ee 
where its precise dependence on the group element $b_{\pm}$ ensures the conservation law $\d_\alpha J^{\alpha}_\pm +[A^{\pm}_{\alpha}, J^{\alpha}_\pm]=0$. The  functional dependence of $\mu$ is chosen so that \eqref{FJ} describes the boundary perturbation of the previous section. Using \eqref{mu}, the radial component reads
\be
J^r_\pm =\frac{K}{2\pi}\delta(t) \Big[V_L\, \delta(x-x_L)-V_R\,\delta(x-x_R)\Big]\,b_{\pm}^{-1}L_{0}b_{\pm}  \,.
\ee
The temporal localization of the external source models a process where the initial and final configurations correspond to solutions of vacuum Einstein's equation.
To construct those solutions in terms of a metric field, we consider a particular radial dependence of the group element $b_{\pm}$,
\be
\label{b}
b_{\pm}(r)=\exp\left[{\pm \frac{r}{2l} (L_1-L_{-1})}\right]\,.
\ee
Since the spatial component of the gauge field \eqref{ax} lies on the hyperbolic conjugacy class,  the resulting gravitational configurations are given by black hole solutions \cite{Afshar:2016wfy,Afshar:2016kjj}. Their geometry close to the Rindler horizon $r=0$ is given by
\be
\label{metrica}
ds^2= \left(\frac{2\pi l}{K}\right)^2( \cI_{\rm tot}^2 dt^2 - 2\cI_{\rm tot} \rho_{\rm tot} dt dx +\rho_{\rm tot}^2 dx^2)+dr^2 + O(r^2) \,, 
\ee
which is expressed in terms of the net electric current $\cI_{\rm tot}=\cI_{+}+\cI_{-}$ and the net charge density has been defined as $\rho_{\rm tot}\equiv\cJ_{+}+\cJ_{-}$. The $O(r^2)$ terms stand for subleading components. One can see that  $\cI_{\rm tot} $ and $\rho_{\rm tot} $ are related to the angular velocity and the horizon area, measured by an observer placed in a rotating frame close to $r=0$.

\subsection{Linear response for near horizon boundary conditions}

We discuss the effect of external source on the spacetime metric \eqref{metrica}, by calculating the  retarded responses of the operators $\cI_{\rm tot}$ and $\rho_{\rm tot}$. The evolution of the metric field is controlled by the equations
\be
\label{eq4.1}
\d_t \cJ_{\pm} + \d_x \cI_{\pm}=\pm \frac{K}{4\pi} \delta(t) \Big[V_L\, \delta(x-x_L)-V_R\,\delta(x-x_R)\Big]\,,
\ee
that has been obtained after using the explicit form of connections \eqref{Aa}. Now, we make use of the boundary condition \eqref{integra}, yielding 
\be
\label{eq4.2}
\cI_\pm=\cI_{\pm}^{\rm in} \pm  v (-)^{\frac{z_\pm+1}{2}} \d_x^{z_\pm-1} \cJ_\pm\,,
\ee
where we have identified $\cI_{\pm}^{\rm in}$ with $a^{\pm}_{0}$. Using \eqref{eq4.1}  and \eqref{eq4.2} subject to initial conditions $\cJ_{\pm}(x,t=-\infty)=\cJ^{\rm in}_{\pm}$, with $\cJ^{\rm in}_\pm$ are constants, the net electric current in frequency space is then given by 
\be
\delta \tilde{\cI}_{\rm tot}=  \int dx' \, \tilde{\mu}(x',\omega)  (\tilde{\chi}^{+}_{\cI,\cJ}+\tilde{\chi}^{-}_{\cI,\cJ})\,,
\ee
where the susceptibilities $\tilde{\chi}^{\pm}_{\cI,\cJ}$ are shown to coincide with the boundary computation \eqref{sss}. This is not a surprise since retarded Green functions play the role of susceptibilities  in the linear response formalism. 

From the bulk point view,  the DC response of the total charge $\rho_{\rm tot} $ can be better understood after Fourier transforming  the equations of motion \eqref{eq4.1}, giving
\be
\label{fourier.cons}
i\omega \tilde{\rho}_{\rm tot} -\d_x \tilde{\cI}_{\rm tot}=0\,.
\ee
In the DC limit, $\delta\tilde{\cI}_{\rm tot}=\sigma (V_L-V_R)+ O(\omega^{\frac{1}{z_\pm}})$ which implies that $\delta \tilde{\rho}_{\rm tot}$
diverges as $O(\omega^{\frac{1}{z_\pm}-1})$ when $\omega \to 0$ \footnote{The case $z_\pm=1$ is excluded from this observation, because it converges. Nonetheless, a more general statement about the observable $\delta\tilde{\rho}_{\rm tot}$ can be drawn from the definition of the area operator \eqref{A}.}. Therefore, generically there is no well-defined DC response for the variation of the total electric charge. However, the  operator
\be
\label{A}
\cA \equiv \int dx \, \rho_{\rm tot}\,,
\ee
representing the horizon area, possesses a more natural interpretation in the context of linear response. In fact, by approximating the integration domain of \eqref{A} to the whole real line, one finds that 
\be
\delta \cA=0\,,
\ee
where we have used \eqref{fourier.cons} and the fact that the electric current vanishes at the boundaries $x \to \pm \infty$, as indicated by \eqref{Ipm}.
From the thermodynamical viewpoint, the constancy of $\cA$ implies that the process sourced by the two-terminal potential \eqref{mu} is adiabatic, provided  the entropy is given by Bekenstein-Hawking formula. 

\subsection{Near horizon memory and chiral bosons}

The spacetime metric \eqref{metrica} provides a further interpretation of the role played by the boundary degrees of freedom. Indeed, the evolution process modeled by \eqref{eq4.1} can be understood as a type of memory effect, arising from an improper gauge transformation where chiral bosons are taken as coordinates. The precise form of such transformation renders AdS$_{3}$ spacetime in a particular frame. 

Let us consider a situation where the source $\mu$ is turned off. By virtue of \eqref{eq:charges} and \eqref{eq:charges2}, one finds
\be
\label{eee}
\pm\frac{4\pi}{K}\left( \cJ_{\pm} dx - \cI_{\pm} dt\right)=d\phi_\pm\,,
\ee
allowing to express the spacetime metric as
\be
\label{AdS}
ds^2=-\frac{l^2}{4} \sinh^2\left(\frac{r}{l}\right)(d\phi_++d\phi_-)^2+dr^2+\frac{l^2}{4} \cosh^2\left(\frac{r}{l}\right)(d\phi_+-d\phi_-)^2\,.
\ee
This foliation corresponds to a double-Wick rotated version of global AdS$_{3}$. Furthermore, the diffeomorphism \eqref{eee} makes evident that the degrees of freedom yielding $\cJ_{\pm}$ and $\cI_{\pm}$ belong to the set of large gauge transformations. 

We are now in a position to analyze a process where the spacetime metric evolves from an initial to a final configuration, both of which satisfy Einstein's equation in vacuum. We start by setting an initial state characterized by
\be
\label{phiin}
\phi^{\rm in}_{\pm}(x,t)=\frac{4\pi}{K} (t \pm x)\cJ^{\rm in}_{\pm}\,,
\ee
representing a BTZ black hole with inner and outer horizons respectively denoted by $\frac{2\pi l^2}{K} (\cJ^{\rm in}_{+} \pm \cJ^{\rm in}_{-})$ (see e.g. \cite{Banados:1998ta}). In order to make this choice consistent with \eqref{eq4.2}, we set the initial value of the electric current to be $\cI_{\pm}^{\rm in}=\mp\cJ_{\pm}^{\rm in}$.

In presence of the source $\mu$, the configuration modifies according to \eqref{eq4.1}. The net effect is encoded in a new chiral boson $\phi^{\rm out}_{\pm}$  parameterizing the charge density and current after the perturbation $\mu$ has ceased. Since the source is localized in time, for $t>0$, they are given by  
\be
\cJ^{\rm out}_\pm= \pm \frac{K}{4\pi} \d_{x} \phi^{\rm out}_{\pm}\,, \quad \cI^{\rm out}_\pm= \mp \frac{K}{4\pi} \d_{t} \phi^{\rm out}_{\pm}\,.
\ee
The field $\phi^{\rm out}_{\pm}$ can be found after solving the equation \eqref{eq4.1}, obtaining
\be
\label{out}
\phi^{\rm out}_{\pm}(x,t)= \phi^{\rm in}_{\pm}(x,t) \mp V_L\, {\psi}_{z_{\pm}}\left(\pm \frac{x-x_L}{(vz_\pm t)^{1/z_\pm}}\right) \pm V_R\, {\psi}_{z_{\pm}}\left(\pm \frac{x-x_R}{(vz_\pm t)^{1/z_\pm}}\right)\,,
\ee
where ${\psi}_{z}(x)$ is a function admitting the following integral representation 
\be
{\psi}_{z}(x)=\frac{1}{2\pi i} \cP \int^{\infty}_{-\infty} \frac{du}{u} \exp\left[i \left(\frac{u^z}{z} + x u \right)\right]\,,
\ee
and $\cP$ is the Cauchy principal value\footnote{For $z>1$, this function satisfies $\d_x {\psi}_{z}(x)={\rm Ai}_{z}(x)$ with ${\rm Ai}_{z}$ the $z$-order Airy function, 
$
{\rm Ai}_{z}(x)=\frac{1}{2\pi} \int^{\infty}_{-\infty} du \exp\left[i \left(\frac{u^z}{z} + x\, u \right)\right]\,
$. See e.g. \cite{Durugo} for more details on higher-order Airy functions.}. 

The final state is given again by the AdS$_{3}$ geometry \eqref{AdS} but now coordinates $\phi_{\pm}$ contain physically relevant information about the evolution process. From the point of view of the canonical charges, the chiral boson  $\phi^{\rm out}_{\pm}$ is an improper gauge transformation as $\cJ_{\pm}$ undergoes a transition from a constant value to a function of $x$ and $t$. Despite the presence of higher-order Fourier modes in the charges, it can be verified that the area operator \eqref{A} does not change in the process.

Another appealing feature codified in the chiral bosons is that the electric current DC response  can be recovered from \eqref{out}. Indeed, the conductivity, defined as the time-average value of the electric current over a voltage gives
\be
\sigma=-\frac{K}{4\pi \Delta V }\int^{\infty}_{-\infty} dt \, \d_t(\delta \phi_{+}-\delta \phi_{-})= \frac{K}{4\pi}\left[{ \psi}_{z^{+}}(0)+{\psi}_{z^{-}}(0)\right]\,,
\ee
where $\delta \phi_{\pm}=\phi^{\rm out}_{\pm}-\phi^{\rm in}_{\pm}$. The result \eqref{conductivity}  is  recovered after using  ${\psi}_{z}(0)=\tfrac{1}{2z}$.

We have found that chiral bosons act as reparameterization modes on AdS space, modifying the charges and storing  DC information of the boundary electric current. These permanent imprints in the solution space emerge as a three-dimensional version of the 4D gravitational memory, where the passage of a gravitational wave induces a permanent displacement of the detectors \cite{Zeldovich:1974gvh,Strominger:2014pwa}.  

It is interesting to compare the present process to the one studied in \cite{Hawking:2016sgy,Donnay:2018ckb}, where a Schwarzschild black hole is hit by a shockwave. Notably, after the wave has fallen into the black hole, the spacetime metric is deformed by an infinitesimal supertranslation that codifies the information of the perturbation. The situation presented here is analogous to \cite{Hawking:2016sgy,Donnay:2018ckb}. Moreover, the chiral bosons $\phi_{\pm}$ can be seen as the finite version of the supertranslation mode controlling the memory in the bulk for any value of the radial coordinate and not only restricted to the near boundary region.

\section{Concluding remarks} \label{conclusion} 

We have identified the boundary degrees of freedom of a $\mathfrak{so}(2,2)$ Chern-Simons theory characterized by temporal connections that depend on higher order derivatives of the fields \eqref{integra}. This leads to a generalized action principle for two copies of anisotropic chiral bosons with Lifshitz symmetry, including the self-dual action as a particular case \cite{Floreanini:1987as}. Inspired by bosonization, we define an electric charge and study the linear response of the associated net current.  It is found that the conductance in the DC limit is give by \eqref{conductivity}
$$
\sigma= \frac{K}{8\pi}\left( \frac{1}{z_+}+\frac{1}{z_-}\right)\,,
$$
showing an explicit dependence on the level of the theory inherited from the Chern-Simons terms and also, a non-trivial dependence on the dynamical exponents of the Lifshitz scaling from each sector $z_{\pm}$.

This boundary transport phenomena have a reinterpretation in terms of gravitational observables. We construct a near horizon metric subject to the same dynamics and boundary perturbation. The disturbance on the spacetimes manifest in two quantities, the total current density $\cI_{\rm tot} $ and the total charge $\rho_{\rm tot} $. The computation of their retarded responses characterize the DC effect in a twofold manner. First, through the horizon area operator $ \cA$ \eqref{A}, where the variation $\delta \rho_{\rm tot} $ implies
$$
\delta \cA=0\,,
$$
meaning that the process is adiabatic in the thermodynamic sense. On the other hand, the variation $\delta\cI _{\rm tot}$ leads to define a conductivity as the time-average value of the electric current over a voltage, recovering the boundary result \eqref{conductivity}. The conductivity at the bulk becomes then a new observable, one that emerges as a result of a type of memory effect. Here, the source marks the transition between two gravitational configurations evolving from a stationary Rindler horizon to a spacetime endowed with soft charges.

 The process can be understood as a large gauge transformation controlled by the chiral bosons, acting as reparametrization modes acting on a global AdS$_3$ spacetime. Indeed, this means that memory effects are not necessarily the result of the passage of gravitational radiation, as three-dimensional gravity is devoid of gravitational waves. 

An important lesson that can be drawn from this work is the construction of a holographic framework to describe transport properties in 1+1 systems as an effect of 3D gravitational memory. To do so, one starts from three-dimensional gravity with a given set of boundary conditions, providing key ingredients allowing to determine the evolution process. The required information to reconstruct a boundary description is summarized in different components of the gauge fields and sources
\begin{itemize}
\item The asymptotic conditions, expressed in the spatial components $a^{\pm}_{x}$, can be used to construct the symplectic kinetic term of the boundary model. 
\item The specification of the temporal components $a^{\pm}_{t}$ yields the Hamiltonian of the boundary theory.
\item The source triggers the linear response in the boundary, corresponding to the external currents coupled to the Chern-Simons bulk equations. In terms of the memory effects, it probes the transition between initial and the final states through the retarded response function.  
\end{itemize}

Another avenue of investigation comes from the inclusion of more boundaries to capture global properties of black holes. As it has been pointed out in \cite{Henneaux:2019sjx}, to properly describe the boundary dynamics of this kind of spacetimes it is necessary to include two boundaries in the Hamiltonian reduction. When the boundary dynamics is computed for topologies with various boundaries, the zero modes couple to the degrees of freedom on each boundary. 
This setup provides a potential mechanism to intertwine transport phenomena on both boundaries.   

Finally, it is desirable to extend the results of this work to different choices of asymptotic conditions. In particular,  those describing non-linear dynamics at the boundary  \cite{Perez:2016vqo,Grumiller:2019tyl,Ojeda:2020bgz,Cardenas:2021vwo}.




\vspace{1cm}

\paragraph{Acknowledgments} 
We thank Francisco Rojas and Ricardo Troncoso for useful discussions. This research has been supported by FONDECYT grants $11190427$, $11190730$, $1181628$, $1210635$ and the grant ANID
Beca Doctorado Nacional 21182110. K. L. would like to thank the organizers of the workshop ``La parte y el todo'', Afunalhue 2021, for the opportunity to present a preliminar version of this work. H.G. would like to thank the support of ANID PIA Anillo ACT/210100 and Proyecto de cooperaci\'on internacional 2019/13231-7 FAPESP/ANID. 

\appendix
\section{Gravity as a Chern-Simons theory}
\label{Appendix A}
General Relativity can be described in terms of the difference of two Chern-Simons actions \cite{Achucarro:1987vz,Witten:1988hc}
\begin{equation}
\label{Two copies of 2+1 CS}
I=I_{CS}[A^+]-I_{CS}[A^-]\,, \quad I_{CS}[A^{\pm}]=\frac{K}{2\pi}\int \left \langle A^{\pm}, dA^{\pm}+\frac{2}{3}A^{\pm2}\right \rangle\,,
\end{equation}
where the gauge fields $A^{\pm}$ are valued in $\mathfrak{sl}(2,\mathbb{R})$ algebra. Its generators are 
\begin{align}
\begin{split}
L^\pm_{-1}=\begin{pmatrix}
0 & 0\\
1 & 0
\end{pmatrix}\;,\; L^\pm_{0}=\begin{pmatrix}
-1/2 & 0 \\
0 & 1/2
\end{pmatrix},\; L^\pm_1=\begin{pmatrix}
0 & -1\\
0 & 0
\end{pmatrix},
\end{split}
\end{align}
whose commutation relations read
\begin{equation}
\comm{L^\pm_n}{L^\pm_m}=(n-m)L^\pm_{n+m}\,.
\end{equation}
The bilinear invariant form is given by  
\begin{align}
\label{Trace identities}
\mean{L^\pm_1,L^\pm_0}=0 \quad\mbox{and}\quad \mean{L^\pm_0,L^\pm_0}=1/2\,.
\end{align}
The connection to three-dimensional gravity is achieved by the identification
\begin{align}
A^\pm&=\omega\pm\frac{e}{l},
\end{align}
where $e$ is the vielbein and $\omega$ is the spin connection.

In section \eqref{Section3}, we have considered Chern-Simons equations in  presence of an external source. At the level of the action principle,  the equations of motion are obtained from
	\be \label{CS}
	I_{CS}[A;J]=I_{CS}[A]-\int d^{3}x  \langle A_{\mu},J^{\mu} \rangle\,. 
	\ee
with
	\be \label{CSgrav}
	I=I_{CS}[A^{+};J_{+}]-I_{CS}[A^{-};J_{-}]\,.
	\ee
Here, $J_\pm$ are  $\mathfrak{sl}(2,\mathbb{R})$ valued currents. Then, the variation of the action \eqref{CSgrav} with respect to the dynamical fields $A^\pm$  becomes 
\be \label{CSpart}
\frac{K}{4\pi}\epsilon^{\mu\nu\rho}F^{\pm}_{\nu\rho}=J_{\pm}^{\mu}\,.
\ee
It is worth noticing that the conservation of the external current $D^{\pm}_\mu J^\mu_\pm=0$ holds on account of the Bianchi identity $\epsilon^{\mu\nu\rho}D^{\pm}_{\mu}F^{\pm}_{\nu\rho}=0$. This fact also ensures gauge invariance of \eqref{CS}.

\section{Bosonization and normal ordering}
\label{Appendix BB}
The interpretation of the charges $\cJ_{\pm}$ as electronic charge densities goes as follows. Following  \cite{Floreanini:1987as}, we define the operators
\be \label{complexfermion}
c_\pm(x)=: e^{-i\sqrt{\frac{K}{2}}\phi_{\pm}(x)}:\quad\text{and}\quad c^{\dagger}_{\pm}(x)=: e^{i\sqrt{\frac{K}{2}}\phi_{\pm}(x)}:\,,
\ee 
where $:\bf{\mathcal{O}}:$ denotes the normal ordering of the operator $\bf{\mathcal{O}}$. It is assumed all the fields are on the same time slice, so the temporal dependence is omitted.  
The commutation relation \eqref{boson comrel} can be cast  as
\begin{align} \label{boson algebra}
\left[ \phi_{\pm} (x),\phi_{\pm}(x') \right]=  \mp \frac{2\pi}{K} i\,{\rm sign}(x-x')\,.
\end{align}
In order to properly deal with the normal order prescription, it is  useful to write down the boson fields $\phi_{\pm}$ as 
\begin{equation} \label{harmonic operators}
\phi_{\pm}=\theta_\pm +\theta^{\dagger}_{\pm}\,,
\end{equation}
where $\theta_\pm $ and $\theta^{\dagger}_{\pm}$ satisfy the commutation algebra, 
\begin{equation} \label{algebraosc}
\left[ \theta_{\pm} (x),\theta^{\dagger}_{\pm}(x') \right]= \pm \frac{2}{K} \log(-2\pi i(x-x'+i\eta))\,,
\end{equation}
 and $\eta$ is an infinitesimal positive regulator.  Notice that given the sign difference in the algebra \eqref{algebraosc} for each copy, the creation and annihilation operators are interchanged.  The pair  $(\theta^{\dagger}_{-},\theta_{-})$ acts following the standard convention, namely, as creation and annihilation operators. On the other hand, creation and annihilation operators corresponds to $(\theta_{+},\theta^{\dagger}_{+})$ in the plus sector, respectively.  The normal order rule imposes creation operators to be placed to the left of all annihilation operators in the product. Hence, one should consider normal-ordering the operators as follows 
 \be \label{complexfermion1}
c_+(x)=e^{-i\sqrt{\frac{K}{2}}\theta_{+}(x)} e^{-i\sqrt{\frac{K}{2}}\theta^{\dagger}_{+}(x)}\quad\text{and}\quad c_-(x)=e^{-i\sqrt{\frac{K}{2}}\theta^{\dagger}_{-}(x)} e^{-i\sqrt{\frac{K}{2}}\theta_{-}(x)}.
\ee 
Considering the above and \eqref{boson algebra}, one can show that the fields $c_{\pm}$ satisfy the anticommutation relations 
$\{c_{\pm}(x),c_{\pm}^\dagger (x')\}=\{c_{\pm}(x),c_{\pm}(x')\}=\{c_{\pm}^\dagger(x),c_{\pm}^\dagger (x')\}= 0$ for $x \neq x'$.
Using the latter, one can express $\cJ_{\pm}$ defined in \eqref{eq:charges}, as fermionic occupation number operators
\begin{equation}
\cJ_{+}(x)=\lim_{x \to x'}  \sqrt{\frac{K}{2}} : c_{+}(x)c_{+}^{\dagger}(x'):\quad\text{and}\quad \cJ_{-}(x)=\lim_{x \to x'} \sqrt{\frac{K}{2}} : c_{-}^{\dagger}(x)c_{-}(x'):.
\end{equation}
This justifies calling $\cJ_{\pm}$ an  electric charge density. 
	
\section{Kubo Formula}
\label{Appendix C}
Here we rederive the expectation value of an operator due to a small perturbation. We consider the following Hamiltonian
\be
H=H_0+\lambda \int dx\, \mu(x) B(x)\,,
\ee 
where $\lambda$ is a small dimensionless quantity, $B$ is a time-independent deformation and $\mu$ is an external source. In the Heisenberg picture, operators evolve with the full Hamiltonian $H$
\be
B(t,x)=e^{i H(t-t_0)} B(x) e^{-i H(t-t_0)}\equiv U^{\dagger}(t,t_0) B_{I}(t,x) U(t,t_0)\,,
\ee
where we have expressed the perturbation in terms of the operators $U$ and $B_I$ given by  
\be
U(t,t_0)=e^{iH_0(t-t_0)} e^{-iH(t-t_0)}, \quad B_I(x,t)=e^{iH_0(t-t_0)} B(x)e^{-iH_0(t-t_0)}\,.
\ee
The  subscript $I$ stands for ``interaction picture" which defines the evolution in terms of the free theory. In our setting, these are the operators we control in perturbation theory.

In order to compute the expectation value, we need to find the evolution of the quantum state. Assuming that for $t \leq t_0$ the state is given by $\left| 0 \right. \rangle$, the wave function for $t>t_0$ is dictated by
\be
\left| \Psi \right. \rangle=  e^{-i H(t-t_0)} \left| 0 \right. \rangle = e^{-i H_0(t-t_0)} U \left| 0 \right. \rangle\,.
\ee
Now we consider $  \langle \left. \Psi \right| A \left| \Psi \right. \rangle$ and we compute it to first order in $\lambda$. This can be done by noticing that $U$ satisfies the equation  
\be
i \frac{\d}{\d t} U= \lambda \left(\int d x' \,\mu(t,x') B_{I}(t,x') \right) U\,,
\ee
therefore it can be  perturbatively  solved in terms of $\lambda$, 
\be
U(t,t_0)= 1- i \lambda \int^{t}_{t_0} dt' dx' \mu(t,x') B_{I}(t,x') +O(\lambda^2)\,.
\ee
Then
\be
\langle A \rangle =  \langle A_{I} \rangle_0 + i \lambda \int d x' \int^{t}_{t_0} dt' \mu(t,x)  \langle  [A_{I}(t,x), B_{I} (t',x')] \rangle_0 +O(\lambda^2)\,.
\ee
Placing the initial state in the far past, $t_0 \to -\infty$, dropping $O(\lambda^2)$ terms and setting $\lambda=1$, we get 
\be
\label{Kubo}
\delta \langle A \rangle =  \int dx' \,dt' \,\, \chi_{A B} (t-t'; x,x')  \mu(t',x')\,,
\ee
with the response function $ \chi_{A B}$  given by
\be
\chi_{A B} (t-t'; x,x')= -i\Theta(t-t') \langle [ A_{I}(t,x), B_{I}(t',x')  ] \rangle\,.
\ee
Through out this work we have used \eqref{Kubo} in terms of the frequency. By Taking Fourier transform one finds
\be
\delta \tilde{ \langle A \rangle} =  \int dx'  \, \tilde{\mu}(\omega,x')\, \tilde{\chi}_{A B} (\omega; x,x')\,,
\ee
where the Fourier transform of a function $O$ is defined by $\tilde{O} \equiv \int^{\infty}_{-\infty}\,dt e^{i\omega t} O $.


\end{document}